\documentclass[useAMS,usenatbib,usegraphicx]{mn2e}

\title[Does Cr 236 host a Cepheid calibrator?]{Does Collinder 236 host a Cepheid calibrator?}
\author[Turner et al.]{D.~G. Turner$^1$\thanks{Email: turner@ap.smu.ca}\thanks{Visiting Astronomer, Helen Sawyer Hogg Telescope, University of Toronto}\thanks{Visiting Astronomer, Harvard College Observatory Photographic Plate Stacks}, D. Forbes$^2\dagger$, P.~J.~T. Leonard$^3$, M. Abdel-Sabour Abdel-Latif$^4$, \and D.~J. Majaess$^1$ and L.~N. Berdnikov$^5\ddagger$\\ \\
$^1$Department of Astronomy and Physics, Saint Mary's University, Halifax, Nova Scotia B3H 3C3, Canada \\
$^2$Department of Physics, Sir Wilfred Grenfell College, Memorial University, Corner Brook, Newfoundland A2H 6P9, Canada \\
$^3$ADNET Systems, Inc., 7515 Mission Dr., Suite A100, Lanham, Maryland 20706, U.S.A. \\
$^4$National Research Institute of Astronomy and Geophysics (NRIAG), Box 11242, Helwan, Cairo, Egypt \\
$^5$Sternberg Astronomical Institute, 13 Universitetskij prosp., Moscow 119992, Russia \\
}
\begin{document}
\date{Accepted 2009 May 6. Received 2009 May 5; in original form 2009 March 28}

\pagerange{\pageref{firstpage}--\pageref{lastpage}} \pubyear{2009}

\maketitle

\label{firstpage}

\begin{abstract}
Photoelectric {\it UBV} photometry and star counts are presented for the previously-unstudied open cluster Collinder 236, supplemented by observations for stars near the Cepheid WZ Car. Collinder 236 is typical of groups associated with Cepheids, with an evolutionary age of $(3.4\pm1.1) \times 10^7$ years, but it is $1944 \pm 71$ pc distant, only half the predicted distance to WZ Car. The cluster is reddened by ${\rm E}_{B-V}\simeq0.26$, and has nuclear and coronal radii of $r_n \simeq 2\arcmin$ (1.1 pc) and $R_c \simeq 8\arcmin$ (4.5 pc), respectively. The Cepheid is not a member of Collinder 236 on the basis of location beyond the cluster tidal radius and implied distance, but its space reddening can be established as E$_{B-V} = 0.268 \pm0.006$ s.e. from 5 adjacent stars. Period changes in WZ Car studied with the aid of archival data are revised. The period of WZ Car is increasing, its rate of $+8.27 \pm 0.19$ s yr$^{-1}$ being consistent with a third crossing of the instability strip.
\end{abstract}


\section{Introduction}

The rich $\eta$ Carinae complex contains many unstudied open clusters, as well as a number of Cepheid variables of potential value as calibrators for the extragalactic distance scale. Previous {\it UBV} surveys of the region for Cepheid/association coincidences by \citet{vb82,vb83} and \citet{te93} were not very successful at discovering potential new calibrators, nor was a more detailed study by \citet{te05} of the unstudied cluster Ruprecht 91 with its nearby Cepheids SX and VY Carinae. There are many Cepheids lying in the region, however, and it is important to examine all potential coincidences with unstudied open clusters to establish whether or not they may constitute physical pairs.

\begin{figure}
\begin{center}
\includegraphics[width=7cm]{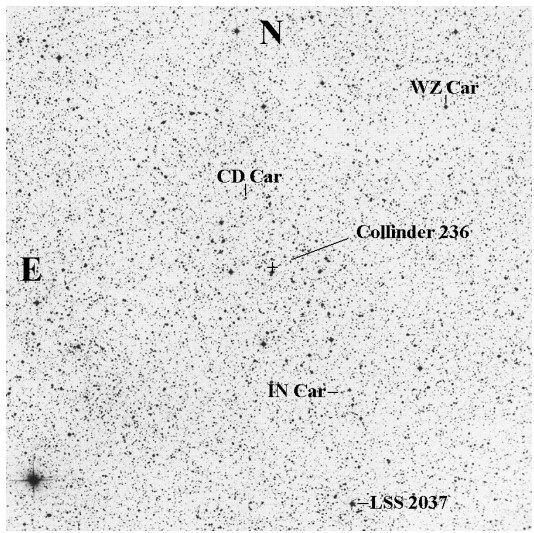}
\end{center}
\caption{The $30\arcmin \times 30\arcmin$ field of Cr 236 from an enlarged portion of the ESO/SRC Atlas of the Southern Sky blue image centred on the adopted cluster centre of symmetry (+ sign), with the location of known variable stars in the field indicated. [The Southern Hemisphere Survey was made with the UK Schmidt Telescope at the Anglo-Australian Observatory, copyright by the UK SERC/PPARC and Royal Observatory Edinburgh.]}
\label{fig1}
\end{figure}

Collinder 236 (C1055--607) is an unstudied, sparsely-populated, open cluster lying southeast of the $\eta$ Carinae Nebula complex. It was noted previously by \citet{ts66} to contain the long-period Cepheid WZ Carinae (HD 94777, $P = 23^{\rm d}$) within its coronal region (see Fig. \ref{fig1}). The Cepheid is roughly 1.5 to 2 magnitudes brighter than the brightest cluster stars, which, in combination with its spatial location, provides sufficient grounds to suspect it may be an outlying cluster member. This paper uses photometric data and star counts for Collinder 236 to examine the likelihood of cluster membership for WZ Car, and to explore its properties.

Galactic star clusters that host Cepheid variables are of more than passing interest, given that they provide a means of calibrating intrinsic colours and luminosities for Cepheids through photometric dereddening procedures and zero-age main sequence (ZAMS) fitting. Recent studies by \citet{bn07} and \citet{vl07} have helped to establish the zero-point of the Cepheid period-luminosity relation through HST and Hipparcos parallaxes, while other calibration techniques have been applied to the problem by \citet{fo07}. There have also been new, deep studies of several recognized Cepheid calibrating clusters by \citet{ho03} and \citet{an07} using CCD {\it UBVK} and {\it BVI$_C$JHK$_5$} photometry, respectively. Such studies are important for extending the calibration sample to long period Cepheids in clusters and associations, although the addition of short-period calibrators is also important, as indicated by the case of CG Cas as a member of Berkeley 58 \citep{te08}. As will be argued elsewhere \citep{tu09}, the open cluster calibration of the Cepheid period-luminosity relation is on fairly solid ground when proper consideration is made for potential problems arising from tie-ins to standard photometric systems, the data reduction techniques employed, the manner in which cluster main-sequence fitting is used to deduce distances, and how interstellar reddening and differential reddening are treated. There remains an immediate need to study new cases as they arise, particularly where they involve long-period Cepheids like WZ Car.

\section{Observations and Star Counts}

Photoelectric {\it UBV} photometry for stars in Cr 236 and near WZ Car was obtained on one night in 1981 May and four nights in 1984 April using the University of Toronto's 0.6-m Helen Sawyer Hogg telescope when it was sited at the Las Campanas Observatory of the Carnegie Institution of Washington. The observations were made in conjunction with similar measurements for stars in Ruprecht 91, as described by \citet{te05}, where details of the data collection and observational uncertainties are given.

\begin{figure}
\begin{center}
\includegraphics[width=7cm]{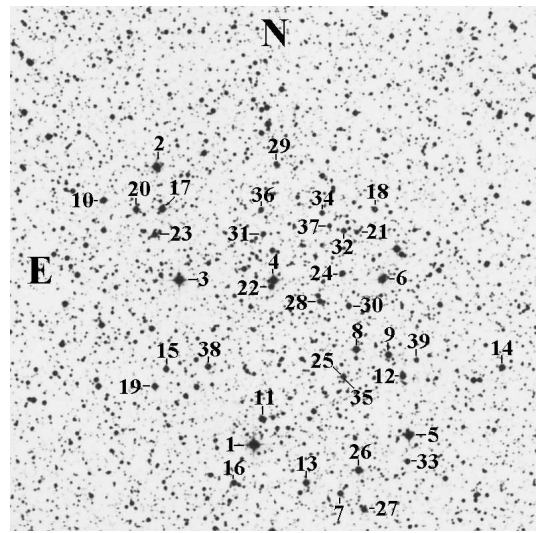}
\end{center}
\caption{A finder chart for the central $15\arcmin \times 15\arcmin$ field of Cr 236 from an enlarged view of Fig. \ref{fig1}, identifying the observed stars.}
\label{fig2}
\end{figure}

Table \ref{tab1} summarizes the observations for stars in Cr 236 (upper section of Table), LSS 2037 and its companion (middle section), and the field of WZ Car (lower section), the latter incorporating the single observations by \citet{vb82}. The {\it V}-magnitude cited for star I by \citet{vb82} appears to have been transcribed incorrectly ({\it V} = 12, not 13), and is corrected here. The brightness of stars 8 and 39 is suspected to vary temporally, but confirmation would require additional photometric measures. Since star 8 is an emission-line B star \citep{he76}, variability would not be unusual. The last column of Table \ref{tab1} indicates the number of nights of observation for each star, and co-ordinates for all stars (epoch 2000.0) are from the 2MASS catalogue \citep{cu03}. Cluster stars are identified in Fig. \ref{fig2}, while stars near WZ Car are identified by the letters and numbers used by \citet{vb82}.

Several luminous stars in the cluster field are identified by \citet{ss71}, including LSS 2040 = Star 6, for which previous {\it UBV} photometry was published by \citet{wr76}. A comparison of the earlier observations with those of Table \ref{tab1} suggests possible small zero-point offsets for the earlier data. The field of the cluster also contains two Algol-type eclipsing binary systems, CD Car and IN Car \citep{sb63,zs85}, but they lie outside the region of the cluster surveyed here. Their locations are indicated in Fig. \ref{fig1}. The M4 Iab type C semi-regular variable CL Car (HD 94599) also lies close to Cr 236, just off the west edge of the field illustrated in Fig. \ref{fig1}. The field immediate east of Cr 236 contains the Of/WN star LSS 2063, which has been studied by \citet{gn87}.

\begin{figure}
\begin{center}
\includegraphics[width=7cm]{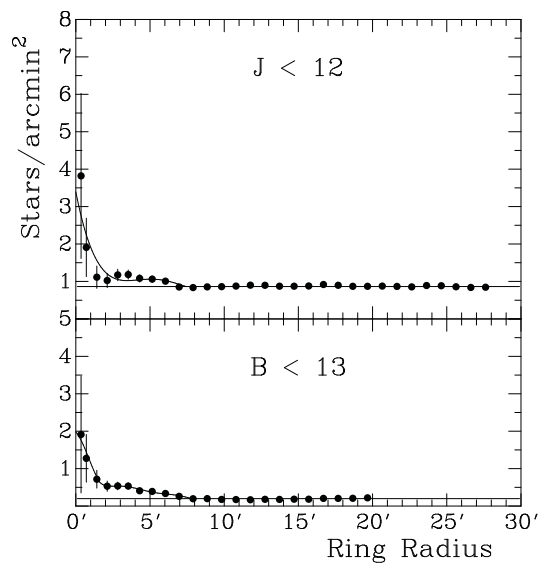}
\end{center}
\caption{Star densities about the adopted centre of Cr 236 for 2MASS data (upper) and counts from the ESO/SRC $B$ image (lower). Best fitting relationships are depicted.}
\label{fig3}
\end{figure}

Star counts were made for Cr 236 using enlargements from the ESO/SRC $B$ southern sky survey transparencies, as done for Ruprecht 91 \citep{te05}. Strip counts were made in several different orientations to delineate the cluster centre, followed by ring counts illustrated in Fig. \ref{fig3}. The latter included counts for stars with $J \le 12$ in the 2MASS catalogue \citep{cu03}. A brightness limit of $B \simeq 13$ was adopted for the ESO/SRC $B$ counts in order to eliminate possible problems arising from background contamination from the rich Carina Nebula complex. The adopted limits in $B$ and $J$ appear to be suitable for investigating density variations across the cluster, which for simplicity was assumed to be spherically symmetric. The cluster centre is located at $10^{\rm h} 56^{\rm m} 57^{\rm s}.1$, $-61^{\circ} 06{\arcmin} 13{\arcsec}$, 2000 co-ordinates, and is illustrated in Fig. \ref{fig1}.

A field star background level of $0.198 \pm 0.012$ stars arcmin$^{-2}$ for $B < 13$ is established from regions lying between $8\arcmin$ and $22\arcmin$ from the cluster centre. Cr 236 is estimated to have a coronal (or tidal) radius of $R_c \simeq 8\arcmin$ and a nuclear radius of $r_n \simeq 2\arcmin$, in the notation of \citet{kh69}, corresponding to 1.1 pc and 4.5 pc, respectively. The number of expected cluster members brighter than $B \simeq 13$ is $33 \pm 7$ on the basis of the Fig. \ref{fig3} star counts. There are only 13 likely cluster members brighter than the counting limit identified within the cluster boundaries in our photometric analysis (with an additional 4 stars lying near WZ Car), so additional cluster members remain to be identified. WZ Car is $15\arcmin.44$ from the centre of Cr 236, beyond the projected cluster tidal radius, so is unlikely to be a cluster member.

The cluster dimensions are somewhat smaller than is typical of other young clusters, although comparable to the similarly small values found for the nearby Carina cluster Ruprecht 91, $r_n = 0.9$ pc and $R_c = 4.3$ pc  \citep{te05}. Perhaps that is indicative of significant loss of member stars over the cluster's lifetime. Several likely members of the cluster main sequence are found beyond the tidal limits, which argues for their recent escape from the main grouping.
 
\setcounter{table}{0}
\begin{table}
\caption[]{Photoelectric {\it UBV} data for stars in Collinder 236.}
\label{tab1}
\begin{center}
\begin{tabular}{@{\extracolsep{-1mm}}rccrrrc}
\hline \noalign{\smallskip}
Star &RA(2000) &DEC(2000) &$V$ &{\it B--V} &{\it U--B} &n \\
\noalign{\smallskip} \hline \noalign{\smallskip}
1* &10 57 04.49 &--61 11 18.0 &8.79 &0.32 &0.15 &2 \\
2 &10 57 23.93 &--61 03 11.4 &9.49 &1.04 &0.81 &2 \\
3* &10 57 20.15 &--61 06 28.8 &9.49 &0.02 &--0.49 &2 \\
4* &10 56 57.77 &--61 06 38.7 &9.77 &0.09 &0.00 &2 \\
5* &10 56 27.54 &--61 11 17.0 &9.86 &0.20 &0.12 &3 \\
6 &10 56 31.88 &--61 06 47.8 &10.62 &0.04 &--0.60 &1 \\
7 &10 56 44.49 &--61 12 50.3 &10.76 &0.23 &0.16 &2 \\
8* &10 56 38.97 &--61 08 45.1 &10.85 &0.06 &--0.59 &3 \\
9 &10 56 31.28 &--61 08 57.2 &10.86 &0.56 &0.05 &2 \\
10 &10 57 36.96 &--61 04 03.3 &10.97 &0.14 &--0.27 &2 \\
11 &10 57 02.37 &--61 10 34.4 &11.00 &0.97 &0.09 &2 \\
12 &10 56 28.05 &--61 09 34.7 &11.09 &0.75 &0.22 &3 \\
13 &10 56 52.55 &--61 12 28.9 &11.22 &0.54 &0.12 &2 \\
14 &10 56 04.54 &--61 09 29.8 &11.34 &0.13 &--0.27 &2 \\
15 &10 57 24.36 &--61 08 59.9 &11.39 &1.18 &1.12 &1 \\
16 &10 57 09.79 &--61 12 21.8 &11.39 &1.05 &0.71 &2 \\
17 &10 57 23.18 &--61 04 24.9 &11.47 &0.13 &--0.32 &2 \\
18 &10 56 32.77 &--61 04 47.2 &11.61 &0.32 &--0.47 &2 \\
19 &10 57 27.31 &--61 09 27.8 &11.66 &0.20 &0.00 &1 \\
20 &10 57 28.83 &--61 04 21.1 &11.69 &0.09 &--0.49 &2 \\
21 &10 56 37.37 &--61 05 24.2 &11.73 &0.97 &0.56 &1 \\
22 &10 56 58.47 &--61 06 49.5 &11.80 &1.46 &1.70 &1 \\
23 &10 57 25.22 &--61 05 07.8 &11.81 &0.60 &0.11 &1 \\
24 &10 56 41.63 &--61 06 31.1 &11.83 &0.15 &--0.44 &2 \\
25 &10 56 42.31 &--61 09 29.8 &11.90 &0.47 &0.22 &2 \\
26 &10 57 27.31 &--61 09 27.8 &11.94 &0.38 &0.28 &2 \\
27 &10 56 39.06 &--61 13 19.9 &12.03 &0.24 &--0.35 &2 \\
28 &10 56 47.08 &--61 07 19.6 &12.08 &0.18 &--0.43 &2 \\
29 &10 56 55.63 &--61 03 18.6 &12.18 &1.29 &1.62 &1 \\
30 &10 56 40.19 &--61 07 30.3 &12.19 &0.12 &--0.40 &3 \\
31 &10 56 59.55 &--61 05 17.5 &12.22 &0.13 &--0.35 &2 \\
32 &10 56 40.39 &--61 05 20.5 &12.27 &0.12 &--0.37 &1 \\
33 &10 56 28.17 &--61 12 02.3 &12.32 &0.10 &--0.45 &3 \\
34 &10 56 43.77 &--61 05 10.9 &12.59 &1.15 &0.75 &1 \\
35 &10 56 41.92 &--61 09 32.0 &12.64 &0.25 &0.21 &1 \\
36 &10 56 59.81 &--61 04 36.0 &12.65 &0.33 &0.16 &1 \\
37 &10 56 45.40 &--61 04 47.8 &12.76 &0.21 &0.03 &1 \\
38 &10 57 14.42 &--61 08 59.8 &12.87 &0.13 &--0.24 &1 \\
39 &10 56 24.82 &--61 09 08.8 &13.01 &0.40 &0.15 &2 \\
\noalign{\smallskip}
2037 &10 56 20.46 &--61 22 03.1 &10.76 &0.22 &--0.56 &2 \\
2037s &10 56 19.82 &--61 22 10.1 &12.44 &0.14 &--0.15 &2 \\
\noalign{\smallskip}
A* &10 54 53.85 &--60 58 03.6 &9.26 &1.44 &1.65 &2 \\
B* &10 55 11.39 &--60 51 30.8 &9.84 &0.15 &--0.64 &1 \\
C &10 55 43.80 &--60 53 33.0 &11.20 &0.08 &--0.51 &1 \\
174 &10 55 24.91 &--60 59 48.0 &11.23 &0.47 &0.13 &2 \\
170 &10 55 13.86 &--60 59 10.2 &11.94 &0.06 &--0.45 &2 \\
D &10 55 30.75 &--60 55 38.0 &12.31 &0.17 &--0.23 &3 \\
I &10 55 17.29 &--60 54 56.9 &12.39 &0.15 &--0.30 &3 \\
E &10 55 12.04 &--60 54 53.6 &12.77 &1.14 &0.90 &1 \\
F &10 55 23.46 &--60 56 51.3 &12.84 &1.34 &1.24 &3 \\
G &10 55 08.57 &--60 57 15.9 &13.05 &1.65 &1.88 &1 \\
172 &10 55 25.66 &--60 57 15.1 &13.22 &1.39 &1.67 &3 \\
H &10 55 08.17 &--60 57 24.7 &13.35 &0.27 &--0.29 &1 \\
J &10 55 17.60 &--60 57 08.0 &13.68 &0.59 &0.22 &4 \\
K &10 55 09.20 &--60 55 04.9 &14.77 &0.39 & $\cdots$ &1 \\
\noalign{\smallskip} \hline
\end{tabular}
\end{center}
Notes: 1 = HD 95032, A4 V; 3 = HD 95056, B3/5 III, 4 = HD 95020, A0 V, 6 = LSS 2040, OBh; 8 = CPD$-60\degr$ 2400, Be; A = HDE 305721, K5; B = HDE 305714, B3/5 Ib.
\end{table}

\section{The cluster Collinder 236}

\begin{figure}
\begin{center}
\includegraphics[width=6cm]{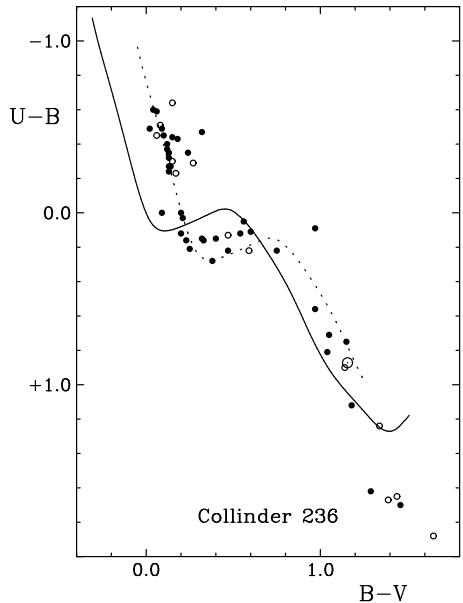}
\end{center}
\caption{A colour-colour diagram for stars in Cr 236 (filled circles) and near WZ Car (open circles). WZ Car is plotted as a circled point. The intrinsic relation for main-sequence stars is plotted as a solid line, and reddened by ${\rm E}_{B-V}=0.26$ as a dotted line.}
\label{fig4}
\end{figure}

Fig. \ref{fig4} presents a {\it UBV} colour-colour diagram for Cr 236 and the field of WZ Car. A distinguishing feature of the plotted colour data is a well-populated sequence of B-type stars reddened by ${\rm E}_{B-V}=0.26$ that represents the luminous end of the cluster main sequence. Several stars at the luminous end of the main sequence appear to be more heavily reddened than the main group, yet there does not appear to be any dependence of the ``extra'' reddening for such stars with spatial location within the cluster. Possibly the stars are affected by a circumstellar reddening component similar to that observed in clusters like Roslund 3 \citep{tu93,tu96a}, although that can only be tested through spectroscopic observation. With the exception of the few anomalous objects, the reddening of cluster stars is remarkably tight, displaying essentially no evidence for differential reddening by dust. Examples of such clusters in the Galactic plane are relatively rare, with the best comparison group to Cr 236 being Stock 16 \citep{tu85}.

\begin{figure}
\begin{center}
\includegraphics[width=7cm]{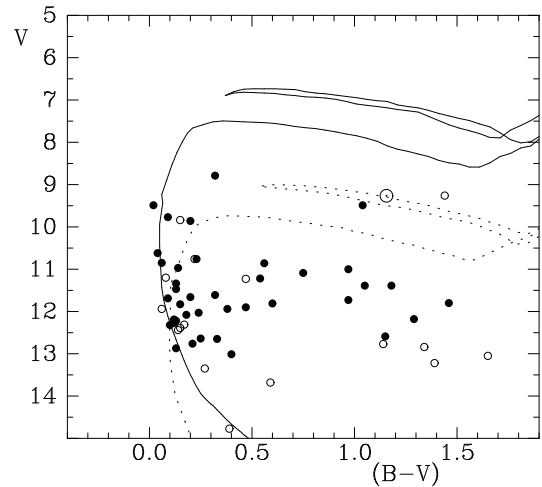}
\end{center}
\caption{A colour-magnitude diagram for stars in Cr 236 (filled circles) and near WZ Car (open circles). WZ Car is plotted as a circled point. An isochrone for $\log t = 7.5$ is plotted for ${\rm E}_{B-V}=0.26$ and $V-M_V=12.44$ as a solid line, and an isochrone for $\log t = 7.6$, ${\rm E}_{B-V}=0.294$, and $V-M_V=14.25$ as a dotted line.}
\label{fig5}
\end{figure}

Fig. \ref{fig5} presents a {\it BV} colour-magnitude diagram for the same stars as in Fig. \ref{fig4}. The evolved B-star main sequence is recognizable on the blue edge of the figure, as well as a possible coeval group of luminous stars discussed later. About half of the stars can be confidently identified as field stars lying along the line of sight to Cr 236 and WZ Car.

An unusual characteristic of the reddening in Carina is that the foreground reddening is described by a shallow reddening slope, ${\rm E}_{U-B}/{\rm E}_{B-V}$ \citep[e.g.,][]{tu89}, and a large ratio of total-to-selective extinction, $R=A_V/{\rm E}_{B-V}=3.82\pm0.13$ \citep{te05}. The sparse spectroscopic data available for Cr 236 stars, from \citet{ss71} and \citet{hc75}, generate colour excesses for Cr 236 stars that imply a reddening law of ${\rm E}_{U-B}/{\rm E}_{B-V}=0.64\pm0.13$ for the field, consistent with the value of ${\rm E}_{U-B}/{\rm E}_{B-V}=0.65\pm0.02$ that applies to the nearby field of Ruprecht 91 \citep{te05}. The latter value, which is tied to a larger sample of stars with spectral types, was adopted here to analyse the photometry for Cr 236 stars.

\begin{figure}
\begin{center}
\includegraphics[width=7cm]{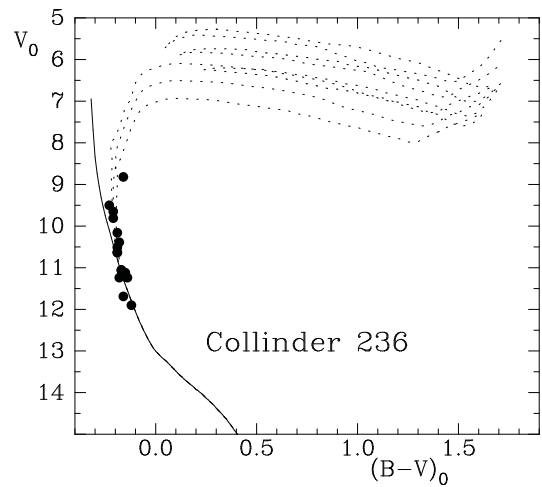}
\end{center}
\caption{Reddening-corrected data for likely members of Cr 236. The ZAMS is plotted as a solid line for $V_0-{\rm M}_V = 11.44$, while dotted lines depict isochrones for $\log t = 7.4$ (upper), $\log t = 7.5$ (middle), and $\log t = 7.6$ (lower).}
\label{fig6}
\end{figure}

Likely members of Cr 236 were corrected for contamination by interstellar extinction using values for the reddening slope ${\rm E}_{U-B}/{\rm E}_{B-V}$ and ratio of total-to-selective extinction $R=A_V/{\rm E}_{B-V}$ found for nearby Ruprecht 91 \citep{te05}. The resulting reddening-free colour-magnitude diagram for the cluster is shown in Fig. \ref{fig6}. Evolution away from the ZAMS appears to affect the data for luminous cluster members, with the implied distance modulus for 5 stars that best represent true ZAMS objects being $V_0-{\rm M}_V = 11.44 \pm 0.08$ s.e., corresponding to a distance of $1944 \pm71$ pc. A comparison with stellar evolutionary model isochrones \citep{me93} for ages of $\log t = 7.4$, 7.5 and 7.6 is shown. All three appear to fit the data for luminous cluster members reasonably well, implying a cluster age of $(3.4\pm1.1) \times 10^7$ years ($\log t = 7.5 \pm0.15$). The corresponding mass of cluster stars falling at the luminous tip of the main-sequence red turnoff (RTO) is $8.5 \pm1.3 M_{\sun}$ \citep{me93}, compared with a value of $M({\rm RTO})\simeq 11 M_{\sun}$ predicted for a cluster containing WZ Car \citep{tu96b}. Although Cr 236 is of similar age to clusters capable of containing long-period Cepheids, its distance is not large enough for it to be associated with WZ Car, which is $\sim 4$ kpc distant according to a reddening-free Cepheid parameterization \citep{me08}.

\begin{figure}
\begin{center}
\includegraphics[width=7cm]{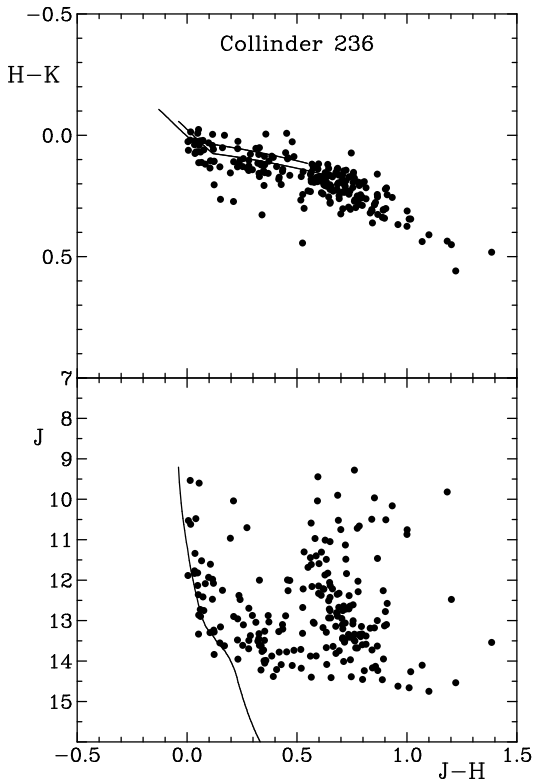}
\end{center}
\caption{A 2MASS colour-colour diagram, {\it H--K} versus {\it J--H} (upper section), for stars within $4\arcmin$ of 10:56:57.1, --61:06:13 (J2000) having magnitude uncertainties less than $\pm0.05$. The intrinsic relation for main-sequence stars is plotted as a solid line, as well as for {\it E(J--H)} $=0.09$. The lower section is a colour-magnitude diagram for the same stars, with a best ZAMS fit of {\it J--M}$_J=11.6$.}
\label{fig7}
\end{figure}

The reality of Collinder 236 was tested using {\it JHK} data from the 2MASS survey \citep{cu03} for stars lying within $4\arcmin$ of the adopted cluster centre and having magnitude uncertainties smaller than $\pm0.05$. The results shown in Fig. \ref{fig7} confirm the existence in the field of a small group of B and A-type stars with similar parameters to those obtained from {\it UBV} photometry. The implied reddening is {\it E(J--H)} $=0.09\pm0.02$, corresponding to {\it E(B--V)} $=0.32\pm0.07$; the distance modulus is $J_0-M_J=11.37\pm0.15$, corresponding to a distance of $1878\pm86$ pc. The results from the 2MASS analysis are tied to a standard interstellar extinction law that relates reddening and extinction in the {\it JHK} bands to those in {\it UBV}, so slight differences from the {\it UBV} analysis can be attributed to the anomalous extinction law applying to foreground extinction in Carina. The results otherwise confirm the reality of Collinder 236 established from the {\it UBV} analysis.

There are a few stars in the region spatially surrounding WZ Car that appear to share a common distance with Cr 236, and hence are possible escaped cluster members, but the Cepheid is not one of them. The case of Cr 236 and WZ Car, like the cluster Ruprecht 91 and the Cepheids VY Car and SX Car, is simply an example of a spatial coincidence between a poorly-populated open cluster and potential calibrating Cepheid.

\setcounter{table}{1}
\begin{table}
\caption{O--C Data for WZ Car.}
\label{tab2}
\begin{tabular}{crcccc}
\hline
JD$_{\rm max}$ &Cycles &O--C &Data &Weight &Source \\
& &(days) &Points \\ 
\hline
2415794.7043 &--1099 &+1.8248 &53 &1.0 &(1) \\
2417381.8949 &--1030 &+1.2219 &67 &1.0 &(1) \\
2418946.3847 &--962 &+0.9297 &70 &1.0 &(1) \\
2420372.8100 &--900 &+0.6420 &70 &1.0 &(1) \\
2421937.7600 &--832 &+0.8100 &41 &1.0 &(1) \\
2423916.0218 &--746 &+0.0828 &24 &1.0 &(1) \\
2428034.9975 &--567 &+0.0000 &42 &1.0 &(1) \\
2428932.4460 &--528 &+0.0000 &70 &1.0 &(1) \\
2429346.6530 &--510 &+0.0000 &70 &1.0 &(1) \\
2430520.0255 &--459 &--0.2140 &70 &1.0 &(1) \\
2432199.7822 &--386 &--0.2968 &70 &1.0 &(1) \\
2433534.5136 &--328 &--0.2324 &32 &1.0 &(1) \\
2434800.4337 &--273 &+0.0552 &16 &2.0 &(2) \\
2441082.5180 &+0 &+0.0000 &40 &3.0 &(3) \\
2441772.6559 &+30 &+0.2071 &13 &3.0 &(4) \\
2444442.8871 &+146 &+0.6901 &56 &3.0 &(5) \\
2448424.4521 &+319 &+1.2656 &113 &2.0 &(6) \\
2450035.3331 &+389 &+1.3416 &50 &3.0 &(7) \\
2450357.6598 &+403 &+1.5073 &27 &3.0 &(8) \\
2450725.9703 &+419 &+1.6338 &50 &3.0 &(7) \\
2451094.0853 &+435 &+1.5648 &50 &3.0 &(7) \\
2451646.5730 &+459 &+1.7765 &40 &3.0 &(9) \\
2451968.7789 &+473 &+1.8214 &22 &3.0 &(10) \\
2452037.8283 &+476 &+1.8363 &70 &2.0 &(11) \\
2452359.9490 &+490 &+1.7960 &42 &3.0 &(12) \\
2452475.0054 &+495 &+1.7949 &70 &2.0 &(11) \\
2452659.1181 &+503 &+1.8156 &37 &3.0 &(13) \\
2452889.3321 &+513 &+1.9146 &70 &2.0 &(11) \\
2453004.3861 &+518 &+1.9111 &24 &3.0 &(14) \\
2453234.6219 &+528 &+2.0319 &70 &2.0 &(11) \\
2453556.9095 &+542 &+2.1585 &70 &2.0 &(11) \\
2453764.1510 &+551 &+2.2965 &70 &2.0 &(11) \\
2454017.1970 &+562 &+2.2160 &70 &2.0 &(11) \\
2454316.5053 &+575 &+2.3748 &70 &2.0 &(11) \\
2454523.5697 &+584 &+2.3357 &70 &2.0 &(11) \\
2454684.7675 &+591 &+2.4530 &70 &2.0 &(11) \\
2454846.0022 &+598 &+2.6072 &37 &2.0 &(11) \\
\hline
\end{tabular}
Data sources: (1) Harvard Collection, (2) \citet{wa58}, (3) \citet{pe76}, (4) \citet{ma75}, (5) \citet{ha80}, \citet{eg83}, and \citet{cc85}, (6) \citet{esa97}, (7) \citet{bt95a,bt95b,bt98,bt00,bt01a}, (8) \citet{br02}, (9) \citet{bc01}, (10) \citet{bt01b}, (11) ASAS \citep{po02}, (12) \citet{bt04b}, (13) \citet{bt04c}, (14) \citet{bt04d}.
\end{table}

\section{The Cepheid WZ Carinae}

The field of WZ Car surveyed here contains 8 stars of B or A spectral types, 5 of which share almost identical reddenings averaging E$_{B-V}{\rm (B0)} = 0.294 \pm0.006$ s.e. The equivalent field reddening for WZ Car is E$_{B-V} = 0.268 \pm0.006$ s.e., compared with  E$_{B-V}\simeq 0.30$ from the reddening parameterization of \citet{me08}. The estimated distance to WZ Car from \citet{me08} is $4-5$ kpc (depending upon which colour relation is adopted), which is much larger than the distance to Cr 236 found here, as noted above.

The Cepheid parameterization of \citet{tu02} leads to an estimate of $\langle M_V\rangle = -4.98$ for WZ Car, and an apparent distance modulus of $V-M_V=14.25$. An isochrone for $\log t = 7.6$ is plotted with those parameters in Fig. \ref{fig5}. Although the surveyed stars cannot be used to establish the distance to WZ Car independently, it is interesting to note the presence in the field of what may be a sequence of luminous stars comparable in age to WZ Car. Star A (HDE 305721) is not among them, despite lying close to the isochrone. Its K5 spectral type indicates that it is an unreddened foreground dwarf. But star B (HDE 305714) is a known B3 Ib supergiant that has an apparent distance modulus of $V-M_V=14.04$ according to the $\beta$-index measured by \citet{gr70}. It appears to be a good match in terms of age and likely distance to the Cepheid. The isochrone fit to WZ Car implies a distance of 4.3 kpc, which suggests that a much deeper survey of the field might be capable of detecting B-type main-sequence stars associated with the Cepheid.

\begin{figure}
\begin{center}
\includegraphics[width=8cm]{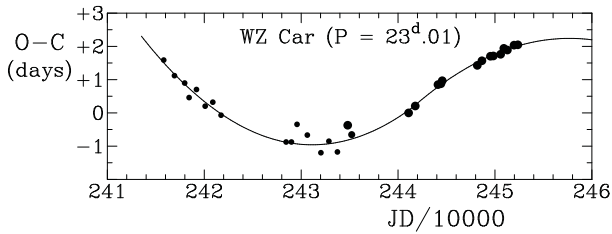}
\end{center}
\caption{O--C diagram for WZ Car based on archival data (small points) and modern photometry (large points). The plotted relationship incorporates two parabolae of opposite sense.}
\label{fig8}
\end{figure}

\begin{figure}
\begin{center}
\includegraphics[width=8cm]{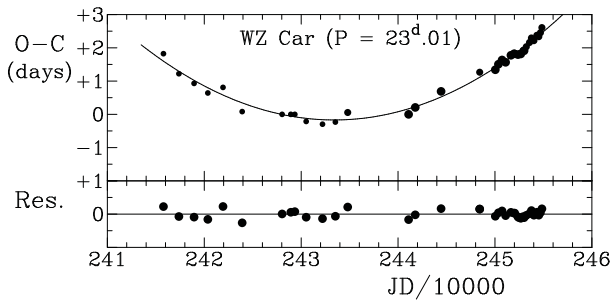}
\end{center}
\caption{Revised O--C diagram for WZ Car based on Table \ref{tab2} data, with point sizes proportional to assigned weight. The bottom diagram plots the O--C residuals from the best fitting parabola, which corresponds to a period increase.}
\label{fig9}
\end{figure}

WZ Car is an intriguing object owing to its period changes \citep{te03}. Although the overall variations from 1900 to the present match a third order polynomial (not a parabola) implying a rate of period increase of $+9.79 \pm 0.30$ s yr$^{-1}$ \citep{bt04a}, recent trends in the O--C data (Fig. \ref{fig8}) appear to suggest a decreasing period, with data prior to 1970 solidly indicating a steadily increasing period. Did the direction of evolution of the Cepheid through the instability strip reverse in 1970?

A reworking of the available data resolves that question. A reanalysis was completed for WZ Car using additonal observational data, improved software, and new light curve templates for {\it B} and {\it V}. The adopted ephemeris for WZ Car for the analysis was:
\begin{displaymath}
{\rm HJD}_{\rm max} = 2441082.518 + 23.0115 \: E ,
\end{displaymath}
with $E$ the number of elapsed cycles. The results are presented in Table \ref{tab2} and plotted in Fig. \ref{fig9}. The observations of \citet{ir61} were omitted from the present analysis, since the phase coverage is poor. Also, only the {\it B}-band observations of \citet{wa58} were used, the {\it V}-band data producing anomalous results. An interesting by-product of the revisions to the O--C data is that they no longer require a third order polynomial to achieve a good fit. A simple parabolic fit works well here.

Previous indications of a changed direction in the evolution of WZ Car are modified significantly by the present analysis. The period is undergoing a continuous increase at $+8.27 \pm 0.19$ s yr$^{-1}$ ($\log {\dot{P}} = +0.917 \pm0.010)$, consistent with a third crossing of the instability strip \citep{te06}. The isochrone for WZ Car plotted in Fig. \ref{fig5} was selected to represent such a stage in the star's evolution.

\section{Discussion}

A photometric analysis of the cluster Cr 236 indicates that it is not associated with the spatially-adjacent long-period Cepheid WZ Car. The cluster appears to be in advanced stages of dissolution into the field, with a number of likely members lying beyond the formal tidal boundaries. Further study of the cluster would be of interest for establishing whether or not the eclipsing systems CD Car and IN Car are likely members, since the known cluster distance would assist in the determination of parameters for the eclipsing components.

Although WZ Car is not a cluster member, it may be part of a loose association of B-type stars in its immediate surroundings. Further study through spectroscopic observation and deeper photometry is needed to assess that possibility. The earlier photographic {\it UBV} survey by \citet{vb82} may not be suitable for such a task, given that the iris measures were calibrated using a comparison sequence tied to an erroneous brightness for one star.

A field reddening for WZ Car can be established from its spatially-adjacent neighbours, the resulting value of E$_{B-V} = 0.268 \pm0.006$ s.e. being consistent with expectations \citep{me08}. Many of the neighbours are only half as distant as the Cepheid, but it appears that there is little or no additional reddening occurring along the line of sight. Potential difficulties regarding the period changes occurring in the pulsation of WZ Car are resolved through a reanalysis of existing data in conjunction with the inclusion of additional observations. WZ Car can now be identified as undergoing a steady period increase consistent with a third crossing of the instability strip.

\subsection*{ACKNOWLEDGEMENTS}
Portions of the present study were supported by research funding awarded through the Natural Sciences and Engineering Research Council of Canada (NSERC), through the Small Research Grants program of the American Astronomical Society, through the Russian Foundation for Basic Research (RFBR), and through the program of Support for Leading Scientific Schools of Russia. We are grateful to the director of Harvard College Observatory for access to the plate stacks, Robert Garrison for the allocation of observing time at UTSO, and Michelle Chouinard Ferguson for software development.

\end{document}